\documentclass[aps,prc,reprint,groupedaddress,onecolumn,floatfix]{revtex4-2}
\usepackage{amsmath}
\usepackage{amssymb}
\usepackage{amsfonts}
\usepackage{amscd}
\usepackage{bm}
\usepackage{eufrak}
\usepackage{graphicx}
\usepackage{color}
\usepackage{hyperref}
\hypersetup{colorlinks=true, citecolor=blue, urlcolor=blue}
\def\beq{\begin{equation}}
\def\eeq{\end{equation}}
\begin{document}

\title{Reflection positivity in Euclidean
formulations of relativistic quantum mechanics of particles.} 

\author{Gohin Shaikh Samad}
\email{ssamad@mcneese.edu}
\affiliation{Department of Mathematical Science\\
McNeese State University,\\
4205 Ryan St., Lake Charles, LA 70605}

\author{W.~N.~Polyzou}
\email{polyzou@uiowa.edu}
\thanks{This work supported by the National Science Foundation,
Award Number \#2427848}
\affiliation{Department of Physics and Astronomy\\ The University of
Iowa\\ Iowa City, IA 52242, USA}

\date{\today}

\begin{abstract}
  
This paper discusses the general structure of reflection positive
Euclidean covariant distributions that can be used to construct
Euclidean representations of relativistic quantum mechanical models of
systems of a finite number of degrees of freedom.  Because quantum
systems of a finite number of degrees of freedom are not local,
reflection positivity is not as restrictive as it is in a local field
theory.  The motivation for the Euclidean approach is that it is
straightforward to construct exactly Poincar\'e invariant quantum
models of finite number of degrees of freedom systems that satisfy
cluster properties and a spectral condition.  In addition the quantum
mechanical inner product can be computed without requiring an analytic
continuation.  Whether these distributions can be generated by a
dynamical principle remains to be determined, but understanding the
general structure of the Euclidean covariant distributions is an
important first step.

\end{abstract}

\maketitle 

\section{Introduction}\label{sec1}

The purpose of this work is to explore a Euclidean formulation of
relativistic quantum mechanics \cite{Kopp:2011vv}
\cite{Polyzou:2013nga},\cite{Polyzou:2019a} \cite{Aiello:2015jgc}
\cite{gohin}.  From a purely academic perspective this Euclidean
formulation yields a Hilbert space structure with a representation of
the Poincar\'e Lie algebra by self-adjoint operators.  The Poincar\'e
generators satisfy cluster properties and the Hamiltonian is bounded
from below.  These elements are sufficient to formulate any kind of
quantum mechanical calculation.  The interesting feature of the
Euclidean representation is that it is not necessary to perform an
analytic continuation to perform these calculations.  The dynamical
input to this representation of relativistic quantum mechanics is a
collection of Euclidean covariant distributions that satisfy a
reflection positivity condition.  Reflection positivity is used to
construct a Hilbert space representation with a non-trivial kernel.
The reflection positivity requirement in the quantum mechanical case
is weaker than the corresponding condition in quantum field theory.
The purpose of this work is to understand the general structure of
these distributions.  What is still missing is a dynamical framework
to generate these model distributions.

The motivation for this exploration is discussed below.  Understanding
the internal structure of elementary hadrons is a goal of
nuclear and particle physics.  Hadronic structure can be studied by
scattering weak or electromagnetic probes off of strongly interacting
hadronic systems.  In a typical reaction the probe scatters off of an
initial hadronic state and causes a transition to a final hadronic
state.  These reactions are complex since the target and strongly
interacting reaction products are composite systems, and particle
number is not generally conserved.  In addition, because the probe can
transfer momentum to the initial hadronic state, the final hadronic
state(s) is in a moving frame relative to the initial state.
Relativistic momentum transfers are needed for a resolution that is
sensitive to the internal structure of hadrons.  QCD is the
appropriate theory to model the hadronic states.  It is a challenging
problem to use QCD to solve for these strongly interacting Poincar\'e
covariant states.  Most non-perturbative calculations utilize
truncations, which are not mathematically controlled approximations.
Lattice calculations are  directly based on QCD, but they break
Poincar\'e symmetry and the treatment of scattering problems with
composite initial and final states is challenging, particularly in
the Euclidean formulation. 

While there are many difficulties associated with solving quantum
field theories, one of them is due to the local nature of quantum
field theories, which cannot be satisfied in theories of a finite
number of degrees of freedom.  The expectation is that for a given
momentum transfer the dynamics will be approximately governed by a
relativistic model of a finite number of relevant degrees of freedom,
limited by the available energy and interaction volume.

Axiomatic formulations of quantum field theory
\cite{Wightman:1980}\cite{Haag:1963dh} \cite{Osterwalder:1973dx}\cite{Osterwalder:1974tc} provide a mathematical
formulation of physical properties that are expected to hold in any
reasonable formulation of relativistic quantum field theory.  The
axioms are internally consistent since they are satisfied by free
field theories.  The challenge is the absence of examples of
non-trivial quantum theories that satisfy the axioms in 3+1
dimensional spacetime.  The property responsible for most of the
difficulties is the locality axiom.  While direct tests of locality on
arbitrarily small distances are not possible, consequences of locality
such as the PCT theorem, crossing symmetry and the need for
antiparticles make it a compelling constraint.

One approach that avoids some of these difficulties is to use
relativistic quantum mechanical models of a finite number of degrees
of freedom to model the initial and final hadronic states
\cite{Bakamjian:1953kh}
\cite{Coester:1965zz}
\cite{Bakker:1979eg}
\cite{Sokolov:1977ym}
\cite{Coester:1982vt}
\cite{Glockle:1986zz}
\cite{Chung:1988mu}
\cite{Chung:1988my}
\cite{Keister:1988zz}
\cite{Polyzou:19892b}
\cite{Chung:1991st}
\cite{Keister:1991sb}
\cite{Cardarelli:1995dc}
\cite{Klink:1996zz}
\cite{Polyzou:1996fb}
\cite{Krutov:1997wu}
\cite{Coester:1997ih}
\cite{Allen:2000ge}
\cite{JuliaDiaz:2003gq}
\cite{Coester:2005cv}
\cite{Lin:2007kg}
\cite{deMelo:2008rj}
\cite{Lin:2008sy}
\cite{Witala:2008va}
\cite{Witala:2009zzb}
\cite{Biernat:2009my}
\cite{Desplanques:2009kj}
\cite{Fuda:2009zz}
\cite{Witala:2011yq}  
\cite{Fuda:2012xd}
\cite{Hadizadeh:2014yua}
\cite{Siraj}
\cite{Grassi}.
These phenomenological models can be constructed so they satisfy most of the
axioms of a quantum field theory, however they are not local, since
they are formulated in terms of a finite number of degrees of freedom.
The advantage is that model states can be calculated using the same
methods that are used in non-relativistic calculations and the
relativistic invariance is exact.  While relativistic models have been
used successfully, their relation to QCD or a more fundamental local
field theory is not direct.
        
While relativistic quantum models do not satisfy microscopic locality,
in order to be useful they need to satisfy cluster properties (
macroscopic locality). This is needed to justify tests of special
relativity on isolated subsystems and more importantly to provide a
connection between few and many-body systems.  While macroscopic
locality can be realized in relativistic models, the construction of
dynamical Poincar\'e generators satisfying macroscopic locality is
based on a recursive construction that uses chains of unitary
transformations that map tensor products of subsystem unitary
representations of the Poincar\'e group and transforms them to a form
where the interactions can be added in a manner that preserves the
Poincar\'e invariance. This is followed by another overall unitary
transformation that restores cluster properties
\cite{Sokolov:1977}\cite{Sokolov:1977ym}
\cite{Coester:1982vt}\cite{Keister:1991sb}.  This is a complicated
construction that has never been used in applications.  Allowing for
production reactions in these models presents an additional set of
problems.

While relativistic quantum models can provide a formal construction of
a relativistically invariant quantum theory satisfying cluster
properties and a spectral condition, it is not a practical solution
for more than two or three-body systems due to the complexity and
non-uniqueness of the construction.  In addition there is no direct
relation to an underlying local relativistic quantum field theory that
can be used to systematically improve the models.  The complexity of
this construction, due to requiring macroscopic locality, and the
absence of a direct connection to a local field theory provides
motivation for exploring alternative non-local formulations of
relativistic quantum mechanics where cluster properties can be
established in a straightforward manner.

This work contributes to an approach based on a Euclidean formulation
of Hamiltonian dynamics \cite{Kopp:2011vv}
\cite{Polyzou:2013nga},\cite{Polyzou:2019a} \cite{Aiello:2015jgc}.  In
the axioms of Euclidean quantum field theory \cite{Osterwalder:1973dx}
\cite{Osterwalder:1974tc} the locality axiom is logically independent
of the other axioms.  This means that it should be possible to
formulate relativistic quantum mechanical models satisfying all of the
other Euclidean axioms.  The theory is expressed in terms of a
collection of Euclidean covariant distributions that satisfy a
condition called reflection positivity \cite{Osterwalder:1973dx}
\cite{Osterwalder:1974tc}\cite{glimm}.  The motivation for considering
this approach is that cluster properties are easy to satisfy.  In
addition, because of the reflection positivity, physical Hilbert space
inner products can be computed without analytic continuation.  The
consequence of not requiring locality is that there is no required
relation between N-point functions with different numbers of initial
and final coordinates.  This makes the positivity of the Hilbert space
norm into a more manageable problem.  It has the additional advantage
that the N-point distributions of a local quantum theory satisfy all
of the required conditions, so there is a clear relation to the
Schwinger functions of a local field theory.

While Wightman functions of a local quantum field theory
\cite{Wightman:1980} also have some of the same advantages, they are
not as closely tied to the dynamics.  Euclidean $N$-point functions
can be approximated in lattice calculations and they are solutions of
the infinite hierarchy of Euclidean Schwinger-Dyson equations.
Scattering is normally formulated in terms of time-ordered Green
functions, which can be obtained from the Euclidean $N$-point
functions by analytic continuation \cite{schwinger}.  The relation to
time-ordered distributions focuses on the scattering matrix rather
than on the Hamiltonian as a Hilbert space operator
\cite{bogoliubov}\cite{Epstein:1973gw}
\cite{grange}\cite{Scharf:2001bk}.  An alternative analytic
continuation gives the Wightman distributions, which are kernels of
the physical Hilbert space inner product.  When the Euclidean
$N$-point distributions are reflection positive, the quantum
mechanical inner product on a dense set of states can be constructed
from the Euclidean distributions without analytic continuation.  There
is a representation of the Poincar\'e Lie algebra
with self-adjoint operators on this
Hilbert space satisfying both a spectral condition and cluster
properties.  These conditions also hold in the non-local case.  It is
even possible to directly formulate time-dependent scattering calculations
in this
representation of the Hilbert space without analytic continuation
\cite{Kopp:2011vv} \cite{Polyzou:2013nga}\cite{Polyzou:2019a}
\cite{Aiello:2015jgc}.

While analytic continuation is needed to construct the Wightman
distributions from the Euclidean distributions, it is not needed to
formulate quantum mechanical calculations. The elements needed for a
relativistic quantum mechanical model are a representation of the
physical Hilbert space inner product, a self-adjoint representation of
the Poincar\'e Lie algebra satisfying cluster properties on this
space, and a Hamiltonian satisfying spectral condition.  Reflection
positivity allows these properties to be satisfied without explicit
analytic continuation.

Reflection positivity ensures both the positivity of the physical
Hilbert space norm and that the spectrum of Hamiltonian is bounded
from below \cite{glimm}.  The resulting spectral condition implies the
existence of the analytic condition, however analytic continuation is
not needed to calculate physical Hilbert space inner products.  In the
Euclidean axioms of a local field theory the $N$-point Schwinger
functions are completely symmetric (anti-symmetric).  They serve as
the kernel of a quadratic form related to the Hilbert space inner
product.  Because of the symmetry one N-point distribution appears in
the kernel of the Euclidean distributions with M initial and K final
variables for any $M+K=N$ \cite{Osterwalder:1973dx}
\cite{Osterwalder:1974tc} \cite{glimm}.  By relaxing the
locality condition a single $N$-point distribution can be replaced by
$N-1$ $N$-point distributions with $M$ initial points
and $N-M$ final points.  Because these are no longer required to be
related by locality, the reflection positivity condition is less
restrictive.

Having a Hilbert space representation and a set of self-adjoint
generators of the Poincar\'e group satisfying cluster properties on
this space are all that is needed to perform calculations.  Cluster
properties can be used to formulate scattering asymptotic conditions
\cite{Brenig:1959}\cite{Ruelle:1962}
\cite{Haag:1958vt}\cite{Aiello:2015jgc}. In the Euclidean case the
dynamics is encoded in the collection of Euclidean covariant distributions 
which will be referred to as quasi-Schwinger
functions.  While some aspects of this approach have been previously
discussed \cite{Kopp:2011vv}
\cite{Polyzou:2013nga}\cite{Polyzou:2019a} \cite{Aiello:2015jgc}
\cite{gohin}, the purpose of this paper is to understand the general
structure of the reflection positive $M+K=N$ point distributions which
provide the dynamical content of this formulation of non-local
relativistic quantum mechanics.  This needs to be understood in order
to address the problem of formulating realistic models.

While the purpose of this work is to demonstrate that it is possible
to construct a large class of non-trivial examples of sets of
Euclidean covariant distributions that are consistent with all of the
desired properties, what is still missing is a dynamical principle
that can be used to determine reflection positive quasi-Schwinger
functions from some simple input.  The fact that Schwinger functions
can be approximated using Euclidean lattice theories or
Schwinger-Dyson equations suggests that this is possible, however this
is beyond the scope of this paper, and will be investigated in future
work.

\section{Notation}\label{sec2}

The following notation will be used.  Euclidean four vectors are denoted
with a subscript $e$: 
\beq
x_e = (x_e^0,x_e^1,x_e^2,x_e^3).
\label{not:1}
\eeq
Minkowski 4 vectors are denoted with a subscript $m$: 
\beq
x_m = (x_m^0,x_m^1,x_m^2,x_m^3).
\label{not:2}
\eeq
The signature of the Minkowski metric is $(-+++)$.
The following matrices:
\beq
\sigma_{m\mu}=
\left (
\left (
\begin{array}{cc}
  1&0\\
  0&1\\
\end{array}  
\right ),
\left (
\begin{array}{cc}
  0&1\\
  1&0\\
\end{array}  
\right ),
\left (
\begin{array}{cc}
  0&-i\\
  i&0\\
\end{array}  
\right ),
\left (
\begin{array}{cc}
  1&0\\
  0&-1\\
\end{array}  
\right )
\right )
\label{not:3}
\eeq
\beq
\sigma_{e\mu}=
\left (
\left (
\begin{array}{cc}
  i&0\\
  0&i\\
\end{array}  
\right ),
\left (
\begin{array}{cc}
  0&1\\
  1&0\\
\end{array}  
\right ),
\left (
\begin{array}{cc}
  0&-i\\
  i&0\\
\end{array}  
\right ),
\left (
\begin{array}{cc}
  1&0\\
  0&-1\\
\end{array}  
\right )
\right )
\label{not:4}
\eeq
are used to construct the $2 \times 2$ matrix representations of Euclidean or Minkowski four vectors:
\beq
X_e= \sum_{\mu=0}^3 x_e^{\mu} \sigma_{e\mu} =
\left (
\begin{array}{cc}
 ix_e^0+x_e^3& x_e^1-i x_e^2\\
 x_e^1+i x_e^2&ix_e^0-x_3^3 \\
\end{array}  
\right )
\qquad x_e^{\mu} = \frac{1}{2} \mbox{Tr}(\sigma^{\dagger}_{e\mu}X_e)
\label{not:5}
\eeq
\beq
X_m= \sum_{\mu=0}^3 x_m^{\mu} \sigma_{m\mu} =
\left (
\begin{array}{cc}
 x_m^0+x_m^3& x_m^1-i x_m^2\\
 x_m^1+i x_m^2&x_m^0-x_m^3 \\
\end{array}  
\right )
\qquad
x_m^{\mu} = \frac{1}{2} \mbox{Tr}(\sigma_{m\mu}X_ )
.
\label{not:6}
\eeq
The Euclidean time-reflection operator, $\theta$, which changes the sign of the
Euclidean time, is defined by 
\beq
\theta x_e = (-x_e^0,x_e^1,x_e^2,x_e^3).
\label{not:7}
\eeq
The Euclidean or Minkowski parity operator $\Pi$ is defined by
\beq
\Pi x_e = (x_e^0,-x_e^1,-x_e^2,-x_e^3)
\qquad
\Pi x_m = (x_m^0,-x_m^1,-x_m^2,-x_m^3).
\label{not:8}
\eeq

\section{Widder's theorem}\label{sec3}

The goal of this work is to understand the general structure of
reflection positive distributions.  The simplest prototype of a 
reflection positive kernel is a $k(t)$ satisfying
\beq
\int_0^\infty dt dt' f(t) k(t+t') f(t') \geq 0
\label{wi:1}
\eeq
for continuous functions $f(t)$ with compact
support on $(0,\infty)$.  

The general form of continuous  
$k(t)$ satisfying (\ref{wi:1}) is (see of \cite{Widder:1934})
\beq 
k(t) =  \int \rho (\lambda) d\lambda e^{-t\lambda}  
\label{wi:2}
\eeq
where $\rho (\lambda)$ is non-decreasing and the integral converges for
$0 \leq t \leq 2T$ where support$(f(t)) \in [0,T]$. 

For $t>0$ this can be written as
\beq
k(t) = \int \tilde{\rho} (\lambda)   \frac{e^{-i t p}}{\lambda^2 + p^2} dp d\lambda
\label{wi:3}
\eeq
where $\tilde{\rho}(\lambda) = \rho (\lambda)\frac{\lambda}{\pi}$ and
the restriction on the support of $f(t)$ means that the $p$ integral
can be computed using the residue theorem.  The
relevant observation is that (\ref{wi:3}) has a structure similar to
a Lehmann representation.  In what follows it is shown that all
positive mass, positive energy unitary irreducible representations of
the Poincar\'e group can be constructed using a generalization of this
construction.  Since any unitary representation of the Poincar\'e
group can be decomposed into direct integrals of irreducible
representations, this provides the general structure of reflection
positive distributions.

\section{Relativistic covariance: $SL(2,{\Bbb C})$ and $SU(2)\times SU(2)$}\label{sec4}
The relation between the Lorentz and orthogonal groups in four dimensions
plays a central role in the connection between the Euclidean and
Minkowski representation of relativistic quantum mechanics.

The covering groups of the complex Lorentz and the four dimensional
complex orthogonal groups are isomorphic.  Because of this there are
formal expressions for the Poincar\'e generators as linear
combinations of the generators of the Euclidean group.  These
operators satisfy the commutation relations of the Poincar\'e Lie
algebra, but they are not self-adjoint on the Hilbert space where the
Euclidean generators are self-adjoint.  Reflection positivity leads to 
a Hilbert space inner product where the Poincar\'e generators
constructed from the Euclidean generators are represented by
self-adjoint operators.

The purpose of this section is to review the relation between the
complex Lorentz and complex orthogonal groups.  The starting point is
to note that Euclidean and Minkowski 4-vectors can be represented by
$2\times 2$ matrices.  Minkowski 4-vectors
$x_m$ can be represented by Hermitian matrices
(\ref{not:5}) 
while the Euclidean 4 vectors $x_e$ can be represented by matrices of
the form (\ref{not:6}).

The advantage of these matrix representations is that
the determinants of these matrices are related to the Minkowski and Euclidean
invariant line elements respectively:
\beq
\mbox{det}(X_m) = (x_m^0)^2 - \mathbf{x}_m\cdot \mathbf{x}_m
\qquad 
\mbox{det}(X_e) = - \left ( (x_e^0)^2 + \mathbf{x}_e\cdot \mathbf{x}_e
\right ).
\label{b.3}
\eeq
In both cases the determinants are preserved under linear
transformations of the form 
\beq
X \to X' = AXB^t
\label{b.4}
\eeq
where $A$ and $B$ are both $SL(2,\mathbb{C})$ matrices and $X$ can be
$X_m$ or $X_e$.  In both cases the relevant symmetry group is $SL(2,\mathbb{C}) \times SL(2,\mathbb{C})$.  For arbitrary elements
$(A,B)\in SL(2,\mathbb{C}) \times SL(2,\mathbb{C})$ the transformation
(\ref{b.4}) defines a complex Lorentz or complex orthogonal
transformation that preserves the real invariant line elements.  The
corresponding complex $4\times 4 $ Lorentz and orthogonal matrices are
\beq
\Lambda(A,B)^{\alpha}{}_{\beta} =
{1\over 2} \mbox{Tr} (\sigma_\alpha
A \sigma_{\beta} B^t) 
\label{b.5}
\eeq
\beq
O(A,B)^{\alpha}{}_{\beta}  =
{1\over 2} \mbox{Tr} (\sigma^{\dagger}_{e\alpha}
A \sigma_{e\beta} B^t). 
\label{b.6}
\eeq
For real Lorentz transformations $B=A^*$ and for real orthogonal
transformations $A$ and $B$ are independent $SU(2)$ matrices.
When the left sides of (\ref{b.5}) and (\ref{b.6})
are real,  the $4 \times 4$ 
transformations can also be expressed by taking complex conjugates
\beq
\Lambda(A,B)^{\alpha}{}_{\beta} = {1\over 2} \mbox{Tr} (\sigma^*_\alpha
A^* \sigma_{\beta}^* B^{\dagger})
\label{b.7}
\eeq
\beq
O(A,B)^{\alpha}{}_{\beta} = {1\over 2} \mbox{Tr} (\sigma^{t}_{e\alpha}
A^* \sigma^*_{e\beta} B^{\dagger}).
\label{b.8}
\eeq

There are two kinds of spinors associated with $SL(2,\mathbb{C})
\times SL(2,\mathbb{C})$ that are distinguished by their
$SL(2,\mathbb{C})$ transformation properties
\beq 
\xi^a \to \xi^{a\prime} = \sum_bA_{ab}\xi^b
\label{b.9}
\eeq
\beq
\xi^{\dot{a}} \to \xi^{\dot{a}\prime} =\sum_{\dot{b}}  B_{\dot{a}\dot{b}}\xi^{\dot{b}}.
\label{b.10}
\eeq
For reasons that will be discussed later spinors transforming like
(\ref{b.9}) are called right-handed spinors while spinors transforming
like (\ref{b.10}) are called left-handed spinors.  Dot superscripts
are are used to distinguish left handed spinors from right handed spinors.
For real Lorentz transformations $B_{\dot{a}\dot{b}}= A^*_{{a}{b}}$
while for real orthogonal transformations $A_{ab}$ and
$B_{\dot{a}\dot{b}}$ are independent $SU(2)$ matrices.  In both cases
the two representations are inequivalent; which means that a general
$A$ and $B$ or $A$ and $A^*$ cannot be related by a single constant
similarity transformation:
\beq
\not\exists \qquad M \mbox{ satisfying } MAM^{-1}=A^* \mbox{ or }
MAM^{-1}=B \qquad \forall \, A,B .
\label{b.11}
\eeq

Because they have determinant 1, a general $SL(2,\mathbb{C})$ matrix
can be expressed as the exponential of the dot product of
a complex 3-vector $\mathbf{z}$ with the vector of Pauli matrices:  
\beq
A = \pm e^{\mathbf{z}\cdot \pmb{\sigma}}
\qquad
\det (A) = e^{\mathbf{z}\cdot \mbox{tr} (\pmb{\sigma})}=1 .
\label{b.12}
\eeq
It follows from the representation (\ref{b.12}) and properties of the Pauli matrices that 
\beq
\sigma_2 A^t \sigma_2 = A^{-1} \qquad \sigma_2 B^t \sigma_2 = B^{-1} .
\label{b.13}
\eeq
This means that $\sigma_2$ behaves like a metric tensor with respect to
these spinors.  Dual right- and left-handed spinors
are identified by lower indices:
\beq
\xi_a:= \sum_b(\sigma_2)_{ab} \xi^b 
\qquad
\xi_{\dot{a}} := \sum_{\dot{b}}(\sigma_2)_{ab} \xi^{\dot{b}}.
\label{b.14}
\eeq
The transformation properties of the dual spinors follow from
the transformation properties of the corresponding upper index spinors
\beq
\xi_a \to \xi_a' = \sum_{bc}(\sigma_2)_{ab} A_{bc} \xi^c =
\sum_{bc}(\sigma_2)_{ab} A_{bc}(\sigma_2)_{cd} \xi_d=
\sum_{b} (A^{t})^{-1}_{ab}\xi_b
\label{b.15}
\eeq
\beq
\xi_{\dot{a}} \to \xi_{\dot{a}}' = \sum_{\dot{b}\dot{c}}(\sigma_2)_{\dot{a}\dot{b}} B_{\dot{b}\dot{c}} \xi^{\dot{c}} =
\sum_{\dot{b}\dot{c}} B_{\dot{a}\dot{b}}(\sigma_2)_{\dot{b}\dot{c}} \xi_{\cdot{c}}=
\sum_{\dot{b}} (B^{t})^{-1}_{\dot{a}\dot{b}}\xi_{\dot{b}}.
\label{b.16}
\eeq
The dual spinors define invariant linear functionals on the
corresponding spinors.  This follows since the contraction of a
spinor and a dual spinor of the same type is invariant under
$SL(2,\mathbb{C})$:
\beq
\sum_a \xi^{a\prime} \chi'_a = \sum_{abc}A_{ab}\xi^b (A^{t})^{-1}_{ac} \chi_c =
\sum_{abc}x^a (A^t)_{ab} ((A^t)^{-1})_{bc} \chi_c =\sum_a \xi^a \chi_a 
\label{b.16a}
\eeq
and replacing $A$ by $B$
\beq
\sum_{\dot{a}} \xi^{\dot{a}\prime} \chi'_{\dot{a}} = \sum_{\dot{a}} \xi^{\dot{a}} \chi_{\dot{a}}.
\label{b.17}
\eeq
For Lorentz transformations when $\mathbf{z}= -i \frac{\pmb{\phi }
}{2}$, $A = \pm e^{\mathbf{z}\cdot \pmb{\sigma}}$ represents a
rotation through an angle $\vert \pmb{\phi} \vert$ about the
$\hat{\pmb{\phi}}$ axis, while for $\mathbf{z}= \frac{\pmb{\rho}}{
2}$, $A = \pm e^{\mathbf{z}\cdot \pmb{\sigma}}$ represents a
rotationless Lorentz boost with rapidity $\pmb{\rho}$.  For rotations
$A$ is unitary while for rotationless boosts $A$ is a positive
Hermitian matrix.  Both $(A,B)$ and $(-A,-B)$ correspond to the
same $4\times4$ complex Lorentz or orthogonal transformation.

The $4\times 4$ matrix representation of the complex orthogonal group
can be expressed in a number of different ways
in terms of the $SL(2,\mathbb{C})$ matrices $A$ and $B$ 
using properties of the Pauli matrices
\beq
\sum_\mu  (O(A,B)x_e)^{\mu}  \cdot \sigma_{e\mu}  =
\sum_\mu x_e^{\mu}  \cdot A \sigma_{e\mu} B^t.
\label{b.17a}
\eeq
Taking transposes gives
\beq
\sum_\mu  ( O(A,B)x_e)^{\mu}  \cdot \sigma^t_{e\mu}  =
\sum_\mu x_e^{\mu}  \cdot B \sigma^t_{e\mu} A^t.
\label{b.17b}
\eeq
Multiplying by $\sigma_2$ on the right and left using
$\sigma_2 A \sigma_2 = (A^t)^{-1}$ on (\ref{b.17a}) and (\ref{b.17b})
gives
\beq
\sum_\mu  (O(A,B)x_e)^{\mu}  \cdot \sigma_2 \sigma_{e\mu} \sigma_2  =
\sum_\mu x_e^{\mu}  \cdot (A^t)^{-1} \sigma_2\sigma_{e\mu}\sigma_2 B^{-1} 
\label{b.17c}
\eeq
and
\beq
\sum_\mu  (O(A,B)x_e)^{\mu}  \cdot \sigma_2 \sigma^t_{e\mu} \sigma_2  =
\sum_\mu x_e^{\mu}  \cdot (B^t)^{-1} \sigma_2\sigma^t_{e\mu}\sigma_2 A^{-1} .
\label{b.17d}
\eeq
In all four cases the complex orthogonal matrix, $O(A,B)^{\mu}{}_{\nu}$,
remains unchanged.

A general $SL(2,\mathbb{C})$ matrix has a polar decomposition
of the form  
\beq
A= RP = P'R'
\label{b.18}
\eeq
where
\[
P:= (A^{\dagger} A)^{1/2} \qquad R := A (A^{\dagger} A)^{-1/2}
\]
\beq
P':= (A A^{\dagger})^{1/2} \qquad R' := (A^{\dagger} A^{\dagger})^{-1/2}A
\label{b.19}
\eeq
and $P$ and $P'$ are positive Hermitian matrices and $R$ and $R'$ are
$SU(2)$ matrices.  This implies that any Lorentz transformation can be
expressed as the product of a rotation and a ``rotationless'' boost in
either order.  Since rotations leave a rest four momentum invariant,
both $A$ and $P'$ in (\ref{b.18}) transform the rest four momentum to
the same final four momentum, $p$, but the ``rotationless'' boost,
$P'$, is distinguished by being a positive matrix.  The rotationaless
boost is a function of the 4-velocity, $v:=p/m$.  The notation
$P(p/m)$ is used for the positive $SL(2,\mathbb{C})$ matrix
corresponding to a boost that transforms $p_0:=(m,\mathbf{0})$ to $p$.

\section{Irreducible representations}\label{sec5}

The elementary building blocks of a relativistic theory of particles
are unitary irreducible representations of the Poincar\'e group
\cite{Wigner:1939cj}.  In this section transformation properties of
single-particle states of mass $m$ and spin $s$ are used to construct
equivalent states in the Euclidean representation.  This work is
limited to positive-mass positive-energy irreducible representations.
These representations are also the building blocks of general unitary
representations of the Poincar\'e group, which can be decomposed into
direct integrals of irreducible representations.

In this section two-point quasi-Schwinger functions that describe
particles of mass $m$ and spin $s$ are constructed.  This
construction motivates the structure of multi-point  reflection positive
quasi-Schwinger functions, which are discussed in the following section.

The starting point is the standard quantum mechanical description of a
particle of mass $m$ and spin $s$.  The state of the particle is
determined by a complete set of compatible measurements.  Compatible
quantities that can be measured are a particle's linear momentum, $\mathbf{p}$, and
the projection, $\mu$, of its spin on a fixed axis in a fixed reference frame.
Lorentz boosts can change the momentum to any value and, because the
spin satisfies $SU(2)$ commutation relations, the spin projection can
take on $2s+1$ values in integer steps between $-s$ and $s$.  These
considerations determine the spectrum of the momentum and
spin projection operators.  A Hilbert
space representation for such a particle is the space of square
integrable functions of the eigenvalues of the linear momentum and a
spin projection operator,
\beq
\langle (m,s) \mathbf{p},\mu \vert \psi \rangle,
\label{ir.1}
\eeq
satisfying 
\beq
\sum_{\mu=-s}^s \int_{\mathbb{R}^3}
d\mathbf{p} \vert \langle (m,s) \mathbf{p},\mu \vert \psi \rangle \vert^2 < \infty .
\label{ir.2}
\eeq

Poincar\'e transformation properties of these wave functions follow
from the transformation properties of the basis of simultaneous
eigenstates of mass, spin, linear momentum and spin projection.  For
particles, $m$ and $s$ are fixed.  In the general case the mass
spectrum is no longer discrete and there are additional invariant degeneracy
quantum numbers.  The group representation property implies that a
general Poincar\'e transformation can be expressed as a Lorentz
transformation followed by a spacetime translation.  It follows that
the unitary representation of the Poincar\'e group, $U(\Lambda ,a )$,
can be expressed as
\beq
U(\Lambda ,a ) = U(I,a) U(\Lambda,0) 
\label{ir.3}
\eeq
where $U(I ,a )$ is a four-parameter unitary representation of the group
of spacetime translations and $U(\Lambda ,0)$ is a unitary representation
of $SL(2,\mathbb{C})$.  The notation $\Lambda$ is used
to represent the $SL(2,\mathbb{C})$ matrix or the corresponding
$4\times 4$ Lorentz transformation.  The interpretation should be
clear from the context.

The spacetime translation operator $U(I,a)$ is a multiplication operator
acting on the basis states
\beq
U(I,a)  \vert (m,s) \mathbf{p},\mu \rangle =
\vert (m,s) \mathbf{p},\mu \rangle e^{i p \cdot a}. 
\label{ir.4}
\eeq
For rotations, $\Lambda=R$, on a $\mathbf{0}$ linear momentum eigenstate,
the rotation does not change the momentum; it only transforms the
magnetic quantum number.  This means that the transformed state is a
linear combination of zero momentum spin states:
\[
U(R,0)\vert (m,s) \mathbf{0},\mu \rangle =
\]
\beq
\sum_{\nu=-s}^s \vert (m,s) \mathbf{0},\nu \rangle
\langle s,\nu \vert U(R,0) \vert s ,\mu \rangle =
\sum_{\nu=-s}^s \vert (m,s) \mathbf{0},\nu \rangle
D^s_{\nu\mu}[R]
\label{ir.5}
\eeq
where $D^s_{\nu\mu}[R]$ is the $2s+1$ dimensional unitary representation
of $SU(2)$ in the basis of eigenstates
of the spin projection $\hat{\mathbf{z}}\cdot \mathbf{s}$,
$D^s_{\nu\mu}[R] = \langle s,\mu \vert U(R,0) \vert s , \nu \rangle$.

The matrix, $D^s_{\nu\mu}[R]$,
can be computed explicitly \cite{schwinger}:
\[
D_{\mu,\nu}^{s}[R]=\langle s,\mu\vert U(R,0) \vert s,\nu\rangle  =
\]
\beq
\sum_{k=0}^{s+\mu}\frac{\sqrt{(s+\mu)!(s+\nu)!(s-\mu)!(s-\nu)!}}{k!
  (s+\nu-k)!(s+\mu-k)!(k-\mu-\nu)!}R_{++}^{k}R_{+-}^{s+\nu-k}
R_{-+}^{s+\mu-k}R_{--}^{k-\mu-\nu}
\label{ir.6}
\eeq
where
\beq
R=
\left (
\begin{array}{cc}
R_{++} & R_{+-}\\
R_{-+} & R_{--}
\end{array}
\right ) = e^{-{i\over 2}\bm{\theta}\cdot \bm{\sigma}}=
\sigma_0 \cos ({\theta \over 2})  - i \hat{\bm{\theta}}\cdot \bm{\sigma}
\sin ({\theta \over 2})
\label{ir.7}
\eeq
is a $SU(2)$ matrix.  $D^s_{\mu \nu}[R]$ in (\ref{ir.6}) is a
homogeneous polynomial of degree $2s$ in the $SU(2)$ matrix elements,
$R_{ij}$, with real coefficients, while the matrix elements $R_{ij}$
are entire functions of angles, $\pmb{\theta}$.  This means that
$D^s_{\mu \nu}[R]$ is an entire function of the rotation angles.

The positive matrix, $P(p/m)=(AA^{\dagger})^{1/2}$,
in the polar decomposition of the
$SL(2,\mathbb{C})$ matrix, $A$, represents a rotationless Lorentz boost. 
It can be used to define
states $\vert (m,s) \mathbf{p},\mu \rangle$
with non-zero linear momentum in terms of zero momentum states,
\beq
\vert (m,s) \mathbf{p},\mu \rangle :=
U(P(p/m),0) \vert (m,s) \mathbf{0},\mu \rangle {N(\mathbf{p})}, 
\label{ir.8}
\eeq
where $N(\mathbf{p})$ is a normalization factor that is chosen to
ensure that this transformation is unitary.  This definition means that
the eigenvalue of the magnetic quantum number in the state (\ref{ir.8}) is the
value that would be measured in the particle's rest frame if
it was boosted to its rest frame by the inverse of a rotationless boost.
The spin defined by
(\ref{ir.8}) is called the canonical spin. (Different spins, like
helicity or light-front spin, can be defined by replacing $P(p/m)$
in (\ref{ir.8}) with a different boost, $A(p/m)=:P(p/m) R(p/m)$, where 
$R(p/m)$ is a momentum dependent rotation).

For basis states with a Dirac delta-function normalization 
\beq
\langle (m,s) \mathbf{p}',\mu' 
\vert (m,s) \mathbf{p},\mu \rangle
= \delta (\mathbf{p}'-\mathbf{p}) \delta_{\mu'\mu}
\label{ir.9}
\eeq
the transformation (\ref{ir.8}) is unitary for
\beq
N(\mathbf{p}) = \sqrt{\frac{m}{\omega_m(\mathbf{p})}}, \qquad
\omega_m(\mathbf{p}):=\sqrt{m^2 +\mathbf{p}^2}. 
\label{ir.10}
\eeq
The unitary transformations, (\ref{ir.4}), (\ref{ir.5})
and (\ref{ir.8}) can be combined to construct a unitary
representation of the Poincar\'e group on the single-particle Hilbert space
\[
U(\Lambda,a) \vert (m,s) \mathbf{p},\mu \rangle =
U(I,a) U(\Lambda,0) \vert (m,s) \mathbf{p},\mu \rangle = 
\]
\[
U(I,a)
U(P (\Lambda p/m),0)
U(P^{-1}(\Lambda p/m),0)
U(\Lambda,0)U(P(p/m),0)\vert (m,s) \mathbf{0},\mu \rangle
\sqrt{\frac{m}
{\omega_m(\mathbf{p})}} = 
\]
\[
\sum_{\nu=-s}^s
U(I,a)
U(P (\Lambda p/m),0)
\vert (m,s) \mathbf{0},\nu \rangle \sqrt{\frac{m}
{\omega_m(\mathbf{p})}}
D^s_{\nu\mu} [(P^{-1}(\Lambda p/m))\Lambda (P(p/m)]
= 
\]
\[
\sum_{\nu=-s}^s
U(I,a)
\vert (m,s) \pmb{\Lambda}p ,\nu \rangle
\sqrt{\frac{\omega_m(\pmb{\Lambda}{p})}
{\omega_m(\mathbf{p})}}
D^s_{\nu\mu} [(P^{-1}(\Lambda p/m))\Lambda (P(p/m)] =
\]
\beq 
\sum_{\nu=-s}^s
e^{i \Lambda p \cdot a}
\vert (m,s) \pmb{\Lambda}p ,\nu \rangle
\sqrt{\frac{\omega_m(\pmb{\Lambda}{p})}
{\omega_m(\mathbf{p})}}
D^s_{\nu\mu} [(P^{-1}(\Lambda p/m))\Lambda (P(p/m)]
\label{ir.11}
\eeq
where $R_w(\Lambda,p):= P^{-1}(\Lambda p/m))\Lambda (P(p/m))$
is a canonical-spin Wigner rotation.  The representation
(\ref{ir.11}) is referred to as a Poincar\'e covariant
unitary representation of the Poincar\'e group.

The next step is to transform (\ref{ir.11}) into a Lorentz covariant
representation of the Poincar\'e group.  Because the Wigner functions
(\ref{ir.6}) are entire functions of angles, both the group
representation property and equations for adding angular momenta with
real Clebsch-Gordan coefficients also hold for complex angles - or
equivalently for all $\Lambda \in SL(2,\mathbb{C})$.

Specifically since
\beq
\sum_{\alpha=-s}^sD^s_{\mu\alpha}[R_2]D^s_{\alpha\nu}[R_1]-D^s_{\mu\nu}[R_2R_1] =0
\label{ir.12}
\eeq
and
\[
D^{s_1}_{\mu_1 \nu_1}[R] D^{s_2}_{\mu_2 \nu_2}[R] -
\]
\beq
\sum_{s,\mu,\nu}
<s_1, \mu_1, s_2, \mu_2|s, \mu> D^{s}_{\mu \nu}[R]
<s_1, \nu_1, s_2, \nu_2|s, \mu>=0
\label{ir.13}
\eeq
are entire and vanish for all real angles,
they vanish for complex angles by analyticity (i.e. for $R=e^{-i\frac{1}{2}\pmb{\theta}\cdot \pmb{\sigma}}\to e^{{\mathbf{z}\over 2}\cdot \pmb{\sigma}}$).  The coefficients
$<s_1, \mu_1, s_2, \mu_2|s, \mu>$ in (\ref{ir.13}) are real $SU(2)$ Clebsch-Gordan coefficients.
This means that the $2s+1$ dimensional representation of the
Wigner rotations can be factored into a matrix product of three
$2s+1$ dimensional
Wigner functions of $SL(2,\mathbb{C})$ matrices (see \ref{ir.6}):
\beq
D^s_{\nu\mu} [R_w(\Lambda,p)] =
\sum_{\nu'\nu''}
D^s_{\nu\nu'} [P^{-1}(\Lambda p/m)]
D^s_{\nu'\nu''}[\Lambda]
D^s_{\nu''\mu}[P(p/m)].
\label{ir.14}
\eeq
Using this decomposition equation (\ref{ir.11}) can be rewritten as
\[
U(\Lambda,0) \sum_{\mu'=-s}^s \vert (m,s) \mathbf{p},\mu' \rangle
D^s_{\mu'\mu}[(P^{-1}(p/m)]
\sqrt{\omega_m(\mathbf{p})}
=
\]
\beq
\sum_{\mu',\mu''=-s}^s
\vert (m,s) \pmb{\Lambda}p,\mu' \rangle D^s_{\mu'\mu''}[P^{-1}(\Lambda p/m)]
\sqrt{\omega_m (\pmb{\Lambda}{p})}
D^s_{\mu''\mu}[\Lambda].
\label{ir.15}
\eeq
This leads to the definition of ``Lorentz covariant'' basis states
\beq
\vert (m,s) \mathbf{p},\mu \rangle_{cov} :=
\sum_{\mu'=-s}^s \vert (m,s) \mathbf{p},\mu' \rangle D^s_{\mu'\mu}[P^{-1}(p/m)]
\sqrt{\omega_m(\mathbf{p})}
\label{ir.16}
\eeq
where the spins transform under a $2s+1$ dimensional representation
of $SL(2,\mathbb{C})$:
\beq
U(\Lambda )  \vert (m,s) \mathbf{p},\mu \rangle_{cov} = \sum_{\mu'=-s}^s
\vert (m,s) \pmb{\Lambda}{p},\mu' \rangle_{cov} D^s_{\mu'\mu}[\Lambda]
.
\label{ir.17}
\eeq
Since for $SU(2)$ matrices, $R = (R^{\dagger})^{-1}$, it also follows that 
\beq
\vert (m,s) \mathbf{p},\mu \rangle_{\overline{cov}} :=
\sum_{\mu'=-s}^s \vert (m,s) \mathbf{p},\mu' \rangle
D^s_{\mu'\mu}[P(p/m)]
\sqrt{\omega_m(\mathbf{p})}
\label{ir.18}
\eeq
which transforms like
\beq
U(\Lambda )  \vert (m,s) \mathbf{p},\mu \rangle_{\overline{cov}} = \sum_{\mu'=-s}^s
\vert (m,s) \pmb{\Lambda}{p},\mu' \rangle_{\overline{cov}} D^s_{\mu'\mu}[(\Lambda^{\dagger})^{-1}].
\label{ir.19}
\eeq
Equations (\ref{ir.17}) and (\ref{ir.19}) define inequivalent
(see (\ref{b.11})) 
unitary
representations of the Lorentz group.

In terms of the notation for the transformation properties of
spinors, the spins in  
(\ref{ir.17}) transform like right handed spinors, $\xi^{\mu}$, while the spins
in (\ref{ir.19}) transform like dual left handed spinors,  $\xi_{\dot{\mu}}$.

Since $D^s_{\mu'\mu}[P(p/m)]$ is invertible, both Lorentz covariant
representations are isomorphic to the original Poincar\'e covariant
representation.  While either Lorentz covariant representation can be
used, these inequivalent representations get transformed into each
other under the discrete transformation of space reflection.  This is
because complex conjugation changes the sign of $\sigma_2$ which
results in $X_m$ being reflected about the $x-z$ plane.  This means
that in order to treat space reflections and Lorentz transformations
consistently in the Lorentz covariant representations, both
representations (\ref{ir.16}) and (\ref{ir.18}) must appear either as
tensor products, as they do for four vectors, or direct sums, as they
do for Dirac spinors.

These covariant basis vectors can be used to construct
Lorentz covariant wave functions:
\beq
_{cov}\langle (m,s) \mathbf{p},\mu \vert \psi \rangle 
\label{ir.21}
\eeq
or
\beq
_{\overline{cov}}\langle (m,s) \mathbf{p},\mu \vert \psi \rangle .
\label{ir.22}
\eeq
The Poincar\'e covariant wave functions are 
related to the Lorentz covariant wave functions by  
\beq
\langle (m,s) \mathbf{p},\mu \vert \psi \rangle =
\sum_{\mu'=-s}^s D^s_{\mu\mu'}[(P)(p/m)]
_{cov}\langle (m,s) \mathbf{p},\mu' \vert \psi \rangle
\sqrt{ \frac{1}{\omega_m(\mathbf{p})}}
\label{ir.23}
\eeq
or
\beq
\langle (m,s) \mathbf{p},\mu \vert \psi \rangle =
\sum_{\mu'=-s}^s D^s_{\mu\mu'}[(P^{-1})(p/m)] 
_{\overline{cov}}\langle (m,s) \mathbf{p},\mu' \vert \psi \rangle 
\sqrt{ \frac{1}{\omega_m(\mathbf{p})}}.
\label{ir.24}
\eeq
The Hilbert space inner product in the Lorentz covariant representation 
follows from relations (\ref{ir.23}) and (\ref{ir.24}) and the inner
product in the 
Poincar\'e covariant irreducible representations (\ref{ir.11}):
\[
\langle \phi \vert \psi\rangle  =
\sum_{\mu=-s}^s \int_{\mathbb{R}^3} d\mathbf{p}
\langle \phi \vert (m,s)\vert \mathbf{p},\mu \rangle \langle (m,s)\mathbf{p},\mu\vert \psi\rangle =
\]
\[
\sum_{\mu,\nu=-s}^s \int_{\mathbb{R}^3} d\mathbf{p}
\langle \phi \vert (m,s)\mathbf{p},\mu \rangle_{cov}  
\frac{D^s_{\mu\nu} [P(p/m)P^{\dagger}(p/m)]}{\omega_m(\mathbf{p})}
_{cov} \langle (m,s)\mathbf{p},\nu\vert \psi\rangle = 
\]
\[
\sum_{\mu,\nu=-s}^s \int_{\mathbb{R}^4} d^4{p}
\langle \phi (m,s)\mathbf{p},\mu \rangle_{cov}
2 \delta (m^2 + p^2) \theta (p^0) \times
\]
\beq
D^s_{\mu\nu} [P(p/m) P(p/m))]
_{cov} \langle (m,s)\mathbf{p},\nu\vert \psi\rangle  
\label{ir.25}
\eeq
with a similar expression for the inner product in the representation (\ref{ir.19}): 
\[
\langle \phi \vert \psi\rangle =
\]
\[
\sum_{\mu,\nu=-s}^s \int_{\mathbb{R}^3} d\mathbf{p}
\langle \phi \vert (m,s)\mathbf{p},\mu \rangle_{\overline{cov}}  
\frac{D^s_{\mu\nu} [P^{-1}(p/m)(P^{\dagger}(p/m))^{-1}]}{\omega_m(\mathbf{p})}
_{\overline{cov}} \langle (m,s)\mathbf{p},\nu\vert \psi\rangle = 
\]
\[
\sum_{\mu,\nu=-s}^s \int_{\mathbb{R}^4} d^4{p}
\langle \phi (m,s)\mathbf{p},\mu \rangle_{\overline{cov}}
2m \delta (m^2 + p^2) \theta (p^0)
\times
\]
\beq
D^s_{\mu\nu} [P^{-1}(p/m))P^{-1}(p/m)]
_{\overline{cov}} \langle (m,s)\mathbf{p},\nu\vert \psi\rangle  
\label{ir.26}
\eeq
where (\ref{b.19}), $P=P^{\dagger}$, was used in (\ref{ir.25})-(\ref{ir.26}).
Because $P(p/m)=e^{\frac{1}{2}\pmb{\rho}\cdot \pmb{\sigma}}$, where $\pmb{\rho}$ is the rapidity of the boost, 
it follows that 
\beq
P(p/m)P(p/m) =e^{\pmb{\rho}\cdot \pmb{\sigma}}=
\frac{\sigma_m \cdot p}{m}
\label{ir.27}
\eeq
and
\beq
P^{-1}(p/m) P^{-1}(p/m) = \frac{(\sigma_2\sigma_m^t \sigma_2)\cdot p}{m} =
\frac{\sigma_m \cdot \Pi p}{m}
\label{ir.28}
\eeq
where $\Pi$ is the parity operator.
These are called ``right'' and ``left-handed'' representations because
they differ by a space reflection.
The resulting kernel is independent of the type of boost
used to define the type of spin (\ref{ir.8}) since for a general
$SL(2,\mathbb{C})$ boost, $A(p/m)$, the polar decomposition (\ref{b.18}) gives 
\[
A(p/m) A^{\dagger}(p/m) = P(p/m)  R(p/m) R^{\dagger}(p/m)P^{\dagger}(p/m)
=
\]
\beq
P(p/m) P^{\dagger}(p/m) = P(p/m)P(p/m) = {\sigma_m \cdot p \over m}.
\label{ir.29}
\eeq
Using (\ref{ir.27}) and (\ref{ir.28}) in
(\ref{ir.25}) and (\ref{ir.26})
gives the following expression for the
inner product of right- or left-handed
Lorentz covariant wave functions:
\[
\langle \phi \vert \psi \rangle =
\]
\[
\sum_{\mu,\nu=-s}^s \int_{\mathbb{R}^3} d^3{p}
\langle \phi \vert (m,s)\mathbf{p},\mu \rangle_{cov}
\frac{D^s_{\mu\nu} [\sigma_m \cdot p / m]} 
{ \omega_m(\mathbf{p})}
_{cov} \langle (m,s)\mathbf{p},\nu\vert \psi\rangle =
\]
\beq
\sum_{\mu,\nu=-s}^s \int_{\mathbb{R}^3} d^4{p}
\langle \phi \vert (m,s)\mathbf{p},\mu \rangle_{cov}
2 \delta (m^2 + p^2) \theta (p^0) 
D^s_{\mu\nu} [\sigma_m \cdot p / m]
_{cov} \langle (m,s)\mathbf{p},\nu\vert \psi\rangle  
\label{ir.30}
\eeq
and
\[
\sum_{\mu,\nu=-s}^s \int_{\mathbb{R}^3} d^3{p}
\langle \phi \vert (m,s)\mathbf{p},\mu \rangle_{\overline{cov}}
\frac{D^s_{\mu\nu} [\sigma_m \cdot \Pi p / m]}{ \omega_m(\mathbf{p})}  
_{\overline{cov}} \langle (m,s)\mathbf{p},\nu\vert \psi\rangle =
\]
\beq
\sum_{\mu,\nu=-s}^s \int_{\mathbb{R}^3} d^4{p}
\langle \phi \vert (m,s)\mathbf{p},\mu \rangle_{\overline{cov}}
2 \delta (m^2 + p^2) \theta (p^0) 
D^s_{\mu\nu} [\sigma_m \cdot \Pi p / m]
_{\overline{cov}} \langle (m,s)\mathbf{p},\nu\vert \psi\rangle . 
\label{ir.31}
\eeq
Note that both $D^s_{\mu\nu} [\sigma_m \cdot p / m]$ and
$D^s_{\mu\nu} [\sigma_m \cdot \Pi p / m]$ are positive since each one can
expressed as the square of a Hermitian matrix.

The Fourier transforms of the kernels of these inner products are 
\beq
W_{R}^s{}_{\mu\nu}(x,y) =
\int_{\mathbb{R}^3} \frac{d^3{p}}{(2\pi)^{3}} e^{ip \cdot (x-y)}
\frac{ D^s_{\mu\nu} [\sigma_m \cdot p / m]}{ \omega_m(\mathbf{p})} 
\label{ir.32}
\eeq
and
\beq
W_L^{s}{}_{{\mu}{\nu}}(x,y) =
\int_{\mathbb{R}^3} \frac{d^3{p}}{(2\pi)^{3}} e^{ip \cdot (x-y)}
\frac{D^s_{{\mu}{\nu}} [\sigma_2\sigma_m^t\sigma_2 \cdot p / m]}
{ \omega_m(\mathbf{p})} 
\label{ir.33}
\eeq
where $p^0=\omega_m(\mathbf{p})$ and $R$ and $L$ stand for ``right'' and ``left''.  The Lorentz covariance properties of these
distributions are
\beq
W_R^{s}{}_{\mu\nu}(\Lambda x,\Lambda y) 
= \sum_{\mu'\nu'= -s}^sD^s_{\mu\mu'}[\Lambda] W_R^{s}{}_{\mu' \nu'}( x, y)
D^s_{\nu'\nu}[\Lambda^{\dagger}]
\label{ir.34}
\eeq
and 
\beq
W_L^{{s}}{}_{{\mu}{\nu}}(\Lambda x,\Lambda y) =
\sum_{\mu'\nu'=-s}^s
D^{{s}}_{{\mu}{\mu'}}[\Lambda^{*}] W_L^{{s}}{}_{{\mu}'{\nu}'}(x,y)
D^{{s}}_{{\nu}'{\nu}}[(\Lambda)^{-1}].
\label{ir.35}
\eeq
Note that in terms of the spinor transformation properties of the
matrices in the kernel, $D^s_{\mu\nu} [\sigma \cdot p / m]$ transforms
a left-handed spinor to a right-handed dual spinor while $D^s_{\mu\nu}
[\sigma \cdot \Pi p / m]$ transforms a right-handed spinor to a
left-handed dual spinor.  This is because complex conjugation
transforms right-handed spinors to left-handed spinors and left-handed
spinors to right-handed spinors.  Because of this property of the
Wigner functions, the dotted index notation will not be used, instead
$D^s_{\mu\nu} [\sigma \cdot p / m]$ will be referred to as the
right-handed representation while $D^s_{\mu\nu} [\sigma \cdot \Pi p / m]$
will be referred to as the left-handed representation.  The invariance
of the inner product is due to the momentum dependence in the kernel.

The kernels of these inner products are representations of two-point
Wightman distributions for right- or left-handed particles with of spin
$s$ \cite{Wightman:1980}.  The price paid in order to have wave functions
with spins that transform under finite dimensional representations of
$SL(2,\mathbb{C})$ is that the inner product has a momentum and
spin-dependent kernel.  In the covariant representations (\ref{ir.16})
and (\ref{ir.18}) the mass dependence has been moved from the
Hamiltonian to the kernel of the inner product.  This is necessary since
time translations and rotationless boosts, which are dynamical,
transform trivially in the Lorentz covariant representation.

A general covariant wave function can transform as a product or direct
sum of right- and left-handed spinor representations.  For product
representations the basis states are replaced by
\beq
\langle \phi \vert (m,s)\mathbf{p},\mu \rangle_{cov} \to
\langle \phi \vert (m,s_r,{s}_l)\mathbf{p},\mu_r,{\mu}_l \rangle_{cov}
\label{ir.36}
\eeq
where the inner product of Lorentz covariant
wave functions has the form
\[
\langle \phi \vert \psi \rangle =
\]
\[
\int_{\mathbb{R}^3} d\mathbf{p}
\sum_{\mu_r,\nu_r=-s_r}^{s_r}
\sum_{\mu_l,\nu_l=-s_l}^{s_l}
\langle \phi \vert (m,s_r,{s}_l)\mathbf{p},\mu_r,{\mu}_l \rangle_{cov}
\times
\]
\[
\frac{
D^{s_r}_{\mu_r\nu_r} [\sigma_m \cdot p / m]
D^{{s}_l}_{{\mu}_l{\nu}_l} [\sigma_m \cdot \Pi p / m]
}{\omega_m (\mathbf{p})}  
\langle  (m,s_r,{s}_l)\mathbf{p},\nu_r,{\nu}_l  \vert \psi \rangle_{cov} =
\]
\[
\int_{\mathbb{R}^4} d^4p \sum_{\mu_r,\nu_r=-s_r}^{s_r}
\sum_{\mu_l,\nu_l=-s_l}^{s_l}
\langle \phi \vert (m,s_r,{s}_l)\mathbf{p},\mu_r,{\mu}_l \rangle_{cov}
\times
\]
\beq
2 \delta (m^2 + p^2) \theta (p^0) 
D^{s_r}_{\mu_r\nu_r} [\sigma \cdot p / m]
D^{{s}_l}_{{\mu}_l{\nu}_l} [\sigma \cdot \Pi p / m]
\langle  (m,s_r,{s}_l)\mathbf{p},\nu_r,{\nu}_l  \vert \psi \rangle_{cov}. 
\label{ir.37}
\eeq

The kernel of this inner product
\[
W^{s_r {s}_l}_{\mu_r {\mu}_l \nu_r {\nu}_l}(x,y) =
\]
\beq
\int {d^4p \over (2 \pi)^4}e^{ip\cdot (x-y)}
2 \delta (m^2 + p^2) \theta (p^0) 
D^{s_r}_{\mu_r\nu_r} [\sigma \cdot p / m]
D^{{s}_l}_{{\mu}_l{\nu}_l} [\sigma \cdot \Pi p / m]
\label{ir.38}
\eeq
is a 2-point Wightman distribution.  The notation $(s_r,s_l)$ will be used
to denote the spin representations in (\ref{ir.38}).

The next step is to relate the Lorentz covariant representation of the
Hilbert space inner product given in terms of the Wightman
distributions to a Hilbert space inner product given in terms of
Euclidean covariant distributions. The representation (\ref{ir.38}) of
the Wightman distributions can be related to a reflection positive
Euclidean two-point distribution.

To show this consider the following Euclidean covariant distribution:
\beq
S^{s_r{s}_l}_{\mu_r {\mu}_l;\nu_r {\nu}_l}( x_e,y_e):=
\int
{d^4p_e \over (2\pi)^4}
\frac{D^{s_r}_{\mu_r\nu_r}[p_e \cdot \sigma_e]
D^{{s}_l}_{{\mu}_l{\nu}_l}[\Pi p_e \cdot \sigma_e]
}{p_e^2 + m^2 } e^{i p_e \cdot ( x_e-y_e)}.
\label{ir.39}
\eeq
Because the Wigner functions are polynomials in the components of
$p_e$, the $p_e$ integral in (\ref{ir.39}) will not generally
converge, however this expression represents a distribution, where it
is necessary to perform the $x_e$ and $y_e$ integrals before computing
the $p_e$ integrals.  If the test functions are Schwartz
functions, their Fourier transforms are Schwartz functions.  This
means that the $p_e$ integrals converge as distributions.  For test
functions satisfying the Euclidean time-support condition, it follows
that
\beq
h^{s_r{s}_l}_{\mu_r {\mu}_l;\nu_r {\nu}_l}(\mathbf{x},\mathbf{y}, p^0_e) := 
\int f^*(\theta x_e) e^{ip^0_e x^0_e} dx_e^0 \int g (y_e) e^{-ip_e^0 y_e^0} dy_e^0
D^{s_r}_{\mu_r\nu_r}[p_e \cdot \sigma_e]
D^{{s}_l}_{{\mu}_l{\nu}_l}[\Pi p_e \cdot \sigma_e]
\label{ir.39a}
\eeq
is analytic in the lower-half $p^0_e$ plane, and 
the  $p_e^0$ integral can be computed using the residue theorem,
closing the contour in the lower half plane.
The poles in
the $p^0_e$ integration are at $p^0_e = \pm i \sqrt{\mathbf{p}^2
+m^2}$. The integral over the contour in the
lower half plane gets a contribution from the pole at $-i \sqrt{\mathbf{p}^2
+m^2}$.

This distribution transforms covariantly under the complex orthogonal group
\[
S^{s_r{s}_l}_{\mu_r {\mu}_l;\nu_r {\nu}_l}(O(A,B)x_e,O(A,B)y_e) =
\]
\beq
\sum_{\mu_r' \nu_r'=-s_r}^{s_r} \sum_{{\mu}'_l\dot{\nu}'_l=-s_l}^{s_l} 
D^{s_r}_{\mu_r\mu_r'}[A]
D^{{s}_l}_{{\mu}_l{\mu}_l'}[(B^t)^{-1}]
S^{s_r{s}_l}_{\mu_r' {\mu}_l';\nu_r' {\nu}_l'}(x_e,y_e)
D^{s_r}_{\nu_r'\nu_r}[B^t]
D^{{s}_l}_{{\nu}_l'{\nu}_l}[A^{-1}].
\label{ir.40}
\eeq
Note that as in the $SL(2,\mathbb{C})$ case, $D^{s_r}_{\mu_r\nu_r}[p_e
\cdot \sigma_e]$ maps a right-handed $SU(2)\times SU(2)$ spinor to a
left-handed dual spinor while $D^{{s}_l}_{{\mu}_l{\nu}_l}[\Pi p_e
\cdot \sigma_e]$ maps a left-handed $SU(2)\times SU(2)$ spinor to a
right-handed dual spinor.


Of interest is when the test functions have support for positive
Euclidean time and the Euclidean time on the final test function is
reflected.  Then the Euclidean time difference in the exponent $\theta
x^e_0-y_e^0$ is strictly negative.
In this case
the result of this integration (for $\theta x^0_e-y^0_e < 0$) is 
\[
\sum_{\mu_r,\nu_r=-s_r}^{s_r} \sum_{\mu_l,\nu_l=-s_l}^{s_l}
\int d^4x_ed^4y_e f^{s_r s_l *}_{\mu_r {\mu}_l}(\theta x_e) 
S^{s\dot{s}}_{\mu_r {\mu}_l;\nu_r {\nu}_l}(x_e,y_e)
g^{s_r s_l}_{\nu \dot{\nu}}(y_e) =
\]
\[
\sum_{\mu_r,\nu_r=-s_r}^{s_r} \sum_{\mu_l,\nu_l=-s_l}^{s_l}
\int d\mathbf{p}  f^{s_r s_l *}_{\mu_r {\mu}_l}( x_e) 
d^4x_e  
\frac{e^{-\omega_m (\mathbf{p})x_e^0 +i \mathbf{p} \cdot \mathbf{x}}}
{(2\pi)^{3/2}}
\times
\]
\beq
\frac{
D^{s_r}_{\mu_r,\nu_r}[p_m \cdot \sigma_m]
D^{{s}_l}_{{\mu}_l,{\nu}_l}[\Pi p_m \cdot \sigma_m]}
{2\omega_m (\mathbf{p})}
d^4 y_e
g^{s_r s_l}_{\nu_r {\nu}_l}(y_e)
\frac{e^{-\omega_m (\mathbf{p})y_e^0 -i \mathbf{p} \cdot \mathbf{y}}}
{(2\pi)^{3/2}}
\label{ir.41}
\eeq
where
\[
(-i \omega_m(\mathbf{p}) ,\mathbf{p}) \cdot (\sigma_e ,\pmb{\sigma} )=
(-i \omega_m(\mathbf{p}) ,\mathbf{p}) \cdot (i \sigma_0 ,\pmb{\sigma} ) =
\]
\beq
(\omega_m(\mathbf{p}) ,\mathbf{p}) \cdot (\sigma_0 ,\pmb{\sigma})
= p_m \cdot \sigma_m
\label{ir.42}
\eeq
was used.

By defining the wave functions
\beq
\psi_{\nu_r {\nu}_l}(\mathbf{p}) :=  
\int_{\mathbb{R}^4} d^4 y_e
g_{\nu_r {\nu}_l}(y_e)
\frac{e^{-\omega_m (\mathbf{p})y_e^0 -i \mathbf{p} \cdot \mathbf{y}}}
{(2\pi)^{3/2}}
\label{ir.43}
\eeq
and
\beq
\phi^*_{\mu_r {\mu}_l}(\mathbf{p}) =
\int_{\mathbb{R}^4} d^4x_e  f^*_{\mu_r {\mu}_l}(x_e)
\frac{e^{-\omega_m (\mathbf{p})x_e^0 +i \mathbf{p} \cdot \mathbf{x}}}
{(2\pi)^{3/2}}
\label{ir.44}
\eeq
(\ref{ir.41}) becomes
\beq
\sum_{\mu_r,\nu_r=-s_r}^{s_r} \sum_{\mu_l,\nu_l=-s_l}^{s_l}\int_{\mathbb{R}^3} 
(\phi^{s_r {s}_l}_{\mu_r {\mu}_l}(\mathbf{p}))^*
 \frac{{d \mathbf{p}}
D^{s_r}_{\mu_r\nu_r}[p_m \cdot \sigma_m]
D^{{s}_l}_{{\mu}_l{\nu}_l}[\Pi p_m \cdot \sigma_m]}
{2\omega_m (\mathbf{p})}
\psi^{s_r {s}_l}_{\nu_r {\nu}_l}(\mathbf{p})
\label{ir.45}
\eeq
which is exactly the expression for the inner product in the
Lorentz covariant representation with wave functions
(\ref{ir.37}) 
\beq
(\phi ^{s_r {s}_l}_{\mu_r {\mu}_l}(\mathbf{p}))^*
=
\langle \phi \vert (m,s_r,{s}_l)\mathbf{p},\mu_r,{\mu}_l \rangle_{cov}
\label{ir.46}
\eeq
and
\beq
\psi^{s_r {s}_l}_{\mu_r {\mu}_l}(\mathbf{p}) =
\langle  (m,s_r,{s}_l)\mathbf{p},\nu_r,{\nu}_l  \vert \psi \rangle_{cov}.
\label{ir.47}
\eeq
Since the Wigner functions of $p_m\cdot \sigma$ and $\Pi p_m\cdot \sigma$
that appear in these expressions are squares of Hermitian matrices,
it follows that (\ref{ir.45}) is non-negative. 
When $B \to A^*$ the complex Euclidean covariance
condition (\ref{ir.40}) becomes the Lorentz covariance condition
(\ref{ir.34}-\ref{ir.35}):
\[
S^{s_r{s}_l}_{\mu_r {\mu}_l;\nu_r {\nu}_l}(O(A,A^*)x_e,O(A,A^*)y_e)
\]
\beq
\sum_{\mu_r', \nu_r'=-s_r}^{s_r} \sum_{{\mu}'_l,{\nu}'_l=-s_l}^{s_l} 
D^{s_r}_{\mu_r\mu_r'}[A]
D^{{s}_l}_{{\mu}_l{\mu}_l'}[(A^{\dagger})^{-1}]
S^{s_r{s}_l}_{\mu_r' {\mu}_l';\nu_r' {\nu}_l'}(x_e,y_e)
D^{s_r}_{\nu_r'\nu_r}[A^{\dagger}]
D^{{s}_l}_{{\nu}_l'{\nu}_l}[A^{-1}].
\label{ir.48}
\eeq
This means that this complex subgroup of the complex orthogonal transformations defines a unitary representation of the Poincar\'e group on the Hilbert space defined by the inner product (\ref{ir.42}).

It follows from (\ref{ir.48}) that 
\[
\sum_{\mu_r', \nu_r'=-s_r}^{s_r} \sum_{{\mu}'_l,{\nu}'_l=-s_l}^{s_l} 
\int f^*(\theta x_e,\mu_r',\mu_l' ) S^{s_r{s}_l}_{\mu_r' {\mu}_l';\nu_r' {\nu}_l'}(x_e,y_e) 
f(y_e ,\nu_r, \nu_l) d^4x_ed^4y_e =
\]
\[
\int 
\sum_{\mu_r', \nu_r'=-s_r}^{s_r} \sum_{{\mu}'_l,{\nu}'_l=-s_l}^{s_l} 
f^*(O(A,A^*)\theta x_e,\mu_r',\mu_l' )
D^{s_r}_{\mu_r\mu_r'}[A]
D^{{s}_l}_{{\mu}_l{\mu}_l'}[(A^{\dagger})^{-1}]
S^{s_r{s}_l}_{\mu_r' {\mu}_l';\nu_r' {\nu}_l'}(x_e,y_e)
\times
\]
\beq
D^{s_r}_{\nu_r'\nu_r}[A^{\dagger}]
D^{{s}_l}_{{\nu}_l'{\nu}_l}[A^{-1}]
f(O(A,A^*) y_e ,\nu'_r, \nu'_l) d^4x_ed^4y_e 
\label{ir.49}
\eeq
or that
\beq
f(y_e ,\nu_r, \nu_l) \to 
\sum_{{\nu}'_r,{\nu}'_l=-s_l}^{s_l}
f(O(A,A^*) y_e ,\nu'_r, \nu'_l)
D^{s_r}_{\nu_r'\nu_r}[A^{\dagger}]
D^{{s}_l}_{{\nu}_l'{\nu}_l}[A^{-1}]
\label{ir.50}
\eeq
is a unitary representation of $SL(2,\mathbb{C})$.

In order to understand how the support
conditions work for finite
Lorentz transformations it is useful to
express the Euclidean coordinates in terms of the matrices $X_e$.
$X_e$  can be decomposed into the sum of Hermitian and anti-Hermitian
parts 
\beq
X_e={1 \over 2} (X_e+ X^{\dagger}_e)+ {1 \over 2} (X_e- X^{\dagger}_e)  =
X_h+X_a
\label{ir.51}
\eeq
where
\beq
X_h=X_h^{\dagger} \qquad
X_a=-X_a^{\dagger}.  
\label{ir.52}
\eeq
Under real Lorentz transformations
\beq
A X_e A^{\dagger} = A X_h A^{\dagger} + A X_a A^{\dagger}
\label{ir.53}
\eeq
the Hermitian and anti Hermitian parts transform independently 
\beq
X_h'= A X_h A^{\dagger}
\qquad
X_a' = A X_a A^{\dagger}.
\label{ir.54}
\eeq
For real Euclidean $X_e$, $X_a= i x^0_eI$ so 
\beq
A X_a A^{\dagger} =  A  A^{\dagger}  X_a = i x^0_e A  A^{\dagger}  
\label{ir.55}
\eeq
which means 
$Tr (A X_a A^{\dagger}) = i x^0_e Tr (AA^{\dagger})$, 
or that the sign of the imaginary part of $X_e$ is unchanged, while the
Hermitian part, associated with real Lorentz transformations
transforms independently. 
This means that the sign of the real part of the Euclidean time is preserved
under real Lorentz transformations. 

Since the integrals are over 4-dimensional Euclidean variables 
\beq
\vert \det AX_e A^{\dagger}\vert = \vert \det X_e \vert = \vert \det X_e^{\dagger} \vert
= \vert \det \theta X_e \vert
\label{ir.56}
\eeq
which implies $d^4x_e= d^4 \theta x_e = d^4 x'_e$ where $x'_e$ is the real Lorentz
transformed $x_e$.

The Fourier transforms of these distributions can be computed explicitly (see \cite{bogoliubov}) 
\[
S^{s_r {s}_l}_{\mu_r {\mu}_l;\nu_r {\nu}_l}(x_e,y_e) =
\]
\[
\int \frac{d^4p_e}{(2\pi)^4}
\int  {e^{i p_e \cdot (x_e-y_e)} D^{s_r}_{\mu_r\nu_r}[p_e \cdot \sigma_e]
D^{{s}_l}_{{\mu}_l{\nu}_l}[p_e \cdot \sigma_2\sigma^*_e\sigma_2]
\over p_e^2 + m^2 } =
\]
\beq
{m^2 \over (2 \pi)^2 }
D^{s_r}_{\mu_r\nu_r}[-i \sigma_e \cdot \nabla_{x_e} ]
D^{{s}_l}_{{\mu}_l {\nu}_l}[-i \sigma_2\sigma^*_e\sigma_2
\cdot \nabla_{x_e}]
{K_1(m \vert x_e-y_e \vert )  \over m\vert x_e-y_e \vert} .
\label{ir.57}
\eeq
The Wigner functions are polynomial in the derivatives. While this kernel is singular
as $x_e \to y_e$,
the Euclidean time support condition along with
the Euclidean time reflection ensures that $\vert x_e-y_e \vert$
never vanishes.

\section{Euclidean covariance}\label{sec6}

The spinor transformation properties of the Schwinger functions
in the previous section,(\ref{ir.40}), were determined by requiring
that the Lorentz invariant inner product is recovered when the
Euclidean wave functions have support for positive Euclidean time and
the final Euclidean time is reflected.

The covariance condition for the Schwinger functions follow from
equation (\ref{ir.40}) which can be expressed in the form
\[
S^{s_r{s}_l}_{\mu_r' {\mu}_l';\nu_r' {\nu}_l'}((O(A,B)x_e,(O(A,B)y_e) =
\]
\beq
\sum_{\mu_r', \nu_r'=-s_r}^{s_r} \sum_{{\mu}'_l,{\nu}'_l=-s_l}^{s_l} 
S^{s_r{s}_l}_{\mu_r' {\mu}_l';\nu_r' {\nu}_l'}(x_e,y_e)
D^{s_r}_{\mu_r'\mu_r}[A^t]
D^{s_r}_{\nu_r'\nu_r}[B^t]
D^{{s}_l}_{{\mu}_l'{\mu}_l}[(B)^{-1}]
D^{{s}_l}_{{\nu}_l'{\nu}_l}[A^{-1}].
\label{eco.3}
\eeq
where $A$ and $B$ are independent $SU(2)$ matrices.  It is
straightforward to extend this condition to Euclidean covariant
kernels with more initial and final Euclidean coordinates and spins.
The distributions that replace the Schwinger functions of axiomatic
field theory will be referred to as quasi-Schwinger functions, since
they are not required to satisfy the locality (symmetry) requirement.

In the Euclidean case, when spinors are involved, this kernel is no
longer positive.  This is most easily seen in equation (\ref{ir.39})
where
$
D^{s_r}_{\mu_r \nu_r } [p_e \cdot \sigma_e]
$ and
$
D^{s_l}_{\mu_l \nu_l } [p_e \cdot \Pi  \sigma_e]
$ are not Hermitian for real $p_e$.

\section{Connected Distributions}\label{sec7}

The construction of positive-mass positive-energy irreducible
representations of the Poincar\'e group using Euclidean covariant
distributions was discussed in section \ref{sec5}.  Tensor products of
these distributions describe non-interacting many-particle systems.

The quasi-Schwinger functions for interacting particles have
cluster expansions, which are sums of tensor products of connected
quasi-Schwinger functions.  Since reflection positivity is preserved
under addition and tensor products, reflection positivity follows if the
connected quasi-Schwinger functions are reflection positive.
Connected quasi-Schwinger functions are the building blocks of general
quasi-Schwinger distributions. The purpose of this section is to show that,
in the absence of the symmetry requirement it is straightforward to construct
connected quasi-Schwinger functions satisfying the conditions needed to
for a relativistic quantum theory.

The structure of connected quasi-Schwinger functions is motivated by
the construction of the quasi-Schwinger functions for positive-mass
positive-energy irreducible representations derived in section (\ref{sec5}).
The general structure of a connected quasi-Schwinger function can be
understood by considering the example of a connected four-point
quasi-Schwinger function with two initial and two final coordinates
and right handed spinors.  It is assumed that it has a contribution
from an ``intermediate state'' with mass $\lambda$ and spin $(s,0)$. 
This exhibits the structural elements of a general
connected quasi-Schwinger functions.

The following structure of the connected quasi-Schwinger function is assumed:
\[
S^c_4(x_{e1}, s_1,\mu_1, x_{e2}, s_2, \mu_2 ; y_{e1}, s_1, \nu_1, y_{e2}, s_2, \nu_2) =
\]
\[
\sum_{\mu, \nu=-s}^{s}
\int
{\cal S}_2^{*s_1, s_2:s} ({1 \over 2} (x_{1e}-x_{e2}),p_e)
\langle s_1, \mu_2, s_2, \mu_2 \vert s, \mu \rangle
\times
\]
\[
{D^s_{\mu\nu} (p_e \cdot \sigma_e) \over p_e^2 + m^2}
e^{i p_e \cdot (x_{e1}+x_{e2}-y_{e1}-y_{e2})} d^4 p_e \times
\]
\beq
\langle s_1, \nu_2, s_2, \nu_2 \vert s, \nu \rangle 
{\cal S}_2^{s_1, s_2:s} ({1 \over 2} (y_{e1}-y_{e2}),p_e).
\label{sc4}
\eeq
where
\beq
{\cal S}_2^{s_1 s_2:s} ({1 \over 2} (x_{e1}-x_{e2}),p_e)
\label{sc5}
\eeq
is a connected Euclidean invariant function of
${1 \over 2} (x_{e1}-x_{e2})$ and $p_e$.  It is assumed to vanish
as $(x_{e1}-x_{e2})^2 \to \infty$, but be analytic in $p_e$.
For identical particles the coefficient
\beq
\langle s_1, \nu_2, s_2, \nu_2 \vert s, \nu \rangle 
{\cal S}_2^{s_1 s_2:s} ({1 \over 2} (y_{e1}-y_{e2}),p_e)
\label{sc6}
\eeq
is assumed to be either symmetric or antisymmetric with respect to interchange of
$1 \leftrightarrow 2$.

To show that this expression satisfies the Euclidean covariance condition
(\ref{eco.3}) note 
\beq
S^c_4(O(A,B)x_{e1}, s_1,\mu_1, O(A,B)x_{e2}, s_2, \mu_2 ; O(A,B)y_{e1}, s_1,\nu_1,
O(A,B)y_{e2} ,s_2, \nu_2) =
\label{sc7}
\eeq
\[
\sum_{\mu, \nu=-s}^{s} 
{\cal S}_2^{*s_1 s_2:s} ({1 \over 2}O(A,B) (x_{e1}-x_{e2}),p_e)
\langle s_1, \mu_2, s_2, \mu_2 \vert s, \mu \rangle \times
\]
\[
{D^s_{\mu\nu} (p_e \cdot \sigma_e) \over p_e^2 + m^2}
e^{i p_e \cdot O(A,B) (x_{e1}+x_{e2}-y_{e1}-y_{e2})} d^4 p_e \times
\]
\beq
\langle s_1, \nu_2, s_2, \nu_2 \vert s, \nu \rangle 
{\cal S}_2^{s_1 s_2:s} (O(A,B) ({1 \over 2} (y_{e1}-y_{e2}),p_e). 
\label{sc10}
\eeq
Euclidean invariance of the dot product in the exponent can be used
to move $O(A,B)$ to $p_e$,
\[
=\int \sum_{\mu, \nu=-s}^{s}
{\cal S}_2^{*s_1 s_2:s} ({1 \over 2} O(A,B) (x_{e1}-x_{e2}),p_e)
\langle s_1, \mu_2, s_2 ,\mu_2, \vert s, \mu \rangle
\times
\]
\[
{D^j_{\mu\nu} ( p_e \cdot \sigma_e) \over p_e^2 + m^2}
e^{i (O(A,B)^{-1} p_e) \cdot  (x_{e1}+x_{e2}-y_{e1}-y_{e2})}d^4 p_e
\]
\beq
\langle s_1, \nu_2, s_2, \nu_2 \vert s, \nu \rangle 
{\cal S}_2^{s_1 s_2:s} ({1 \over 2} O(A,B) (y_{e1}-y_{e2}), p_e) 
\label{sc13}
\eeq
while changing variables $p_e'= O^{-1}(A,B)p_e$ gives 
\[
=\int \sum_{\mu, \nu=-s}^{s}
{\cal S}_2^{*s_1 s_2:s} ({1 \over 2} O(A,B) (x_{e1}-x_{e2}),O(A,B)p_e)
\langle s_1, \mu_2, s_2 ,\mu_2, \vert s, \mu \rangle
\times
\]
\[
{D^s_{\mu\nu} (O(A,B)  p_e \cdot \sigma_e) \over p_e^2 + m^2}
e^{i p_e \cdot  (x_{e1}+x_{e2}-y_{e1}-y_{e2})}d^4 p_e
\times
\]
\beq
\langle s_1, \nu_2, s_2, \nu_2 \vert s, \nu \rangle 
{\cal S}_2^{s_1 s_2:s} ({1 \over 2} O(A,B) (y_{e1}-y_{e2}),O(A,B) p_e) .
\label{sc16}
\eeq
The Euclidean invariance,  ${\cal S}_2^{s_1 s_2:s} ({1 \over 2} (y_{e1}-y_{e2}), p_e) =  {\cal S}_2^{s_1 s_2:s} (O(A,B){1 \over 2} (y_{e1}-y_{e2}),O(A,B) p_e)$
means that the factors of $O(A,B)$ can  be removed while (\ref{b.14}), 
$(O(A,B)(p_e \cdot \sigma_e) = A  (p_e \cdot \sigma_e) B^{t}$,
gives
\[
= \sum_{\mu, \nu=-s}^{s}
{\cal S}_2^{*s_1 s_2:s} ({1 \over 2} (x_{e1}-x_{e2}),p_e)
\langle s_1, \mu_1, s_2, \mu_2 \vert s, \mu \rangle
\times
\]
\[
D^s_{\mu\mu'} (A ){D^s_{\mu'\nu'}(  p_e \cdot \sigma_e) \over p_e^2 + m^2}
D^s_{\nu'\nu} (B^{t})  
e^{i p_e \cdot  (x_{e1}+x_{e2}-y_{e1}-y_{e2})} d^4 p_e
\]
\beq
\langle s_1, \nu_1, s_2, \nu_2 \vert s, \nu \rangle 
{\cal S}_2^{s_1 s_2:s} ({1 \over 2} (y_{e1}-y_{e2}), p_e).
\label{sc19}
\eeq
Finally the property (\ref{ir.13}) of the Clebsch-Gordan coefficients can be used to get
\[
=\int \sum_{\mu, \nu=-s}^{s}\sum_{\mu_1', \nu_1'=-s_1}^{s_1}
\sum_{\mu_2', \nu_2'=-s_2}^{s_2} 
D^{s_1}_{\mu_1\mu_1'} (A) D^{s_2}_{\mu_2\mu'_2} (A)
{\cal S}_2^{*s_1 s_2:s} ({1 \over 2} (x_{e1}-x_{e2}),p_e)
\langle s_1, \mu_1', s_2, \mu_2' \vert s, \mu \rangle
\times
\]
\[
{D^s_{\mu\nu} ( p_e \cdot \sigma_e) \over p_e^2 + m^2}
e^{i p_e \cdot  (x_{e1}+x_{e2}-y_{e1}-y_{e2})} d^4 p_e
\]
\beq
\langle s_1, \nu_1', s_2, \nu_2' \vert s, \nu \rangle 
{\cal S}_2^{s_1 s_2:s} ({1 \over 2} (y_{e1}-y_{e2}), p_e)
D^{s_1}_{\nu_2'\nu_2} (B^{t}) D^{s_2}_{\nu_2'\nu_2} (B^{t}) 
\label{sc22}
\eeq
which shows that the  two-point quasi Schwinger function (\ref{sc4})
satisfies the Euclidean covariance condition (\ref{eco.3}):
\beq
=
\sum_{\mu_1', \nu_1'=-s_1}^{s_1}
\sum_{\mu_2', \nu_2'=-s_2}^{s_2} 
D^{s_1}_{\mu_1\mu_1'} (A) D^{s_2}_{\mu_2\mu'_2} (A)
S^c_4(x_{e1}, s_1,\mu_{1}',x_{e2}, s_2, \mu_2' ;y_{e1}, s_1,\nu_1',
y_{e2} ,s_2, \nu_2')
D^{s_1}_{\nu_2'\nu_2} (B^{t}) D^{s_2}_{\nu_2'\nu_2} (B^{t}) 
\label{sc23}
\eeq
 
The other requirement is reflection positivity.  In this case it
is sufficient to assume that the test functions have support for
positive Euclidean times.  It follows that they have support for
positive $X^0_e= {1 \over 2}(x^0_{e1}+x^0_{e2})$. 

To show reflection positivity note that
$(\theta f,S f)$ is:
\beq
\int \sum_{\mu_1, \nu_1=-s_1}^{s_1}
\sum_{\mu_2, \nu_2=-s_2}^{s_2}  f^* (\theta x_{e1}, s_1, \mu_1 , \theta x_{e2},s_2, \mu_2 ) 
S^c_4(x_{e1}, s_1,\mu_1,x_{e2}, s_2, \mu_2 ;y_{e1}, s_1,\nu_1,
y_{e2} ,s_2, \nu_2)
f( y_{e1}, s_1, \nu_1 , y_{e2},s_2 ,\nu_2 ). 
\label{sc24}
\eeq
Moving the reflection operators from the final test functions to the quasi-Schwinger function gives
\[
=\int\sum_{\mu_1, \nu_1=-s_1}^{s_1}
\sum_{\mu_2, \nu_2=-s_2}^{s_2}   f^* ( x_{e1}, s_1, \mu_1 , x_{e2},s_2 ,\mu_2 ) 
{\cal S}_2^{*s_1 s_2:s} ({1 \over 2} (\theta (x_{e1}-x_{e2)},p_e)
\langle s_1, \mu_1, s_2, \mu_2 \vert s, \mu \rangle
\times
\]
\[
{D^s_{\mu\nu} ( p_e \cdot \sigma_e) \over p_e^2 + \lambda^2}
e^{i p_e \cdot  (\theta x_{e1}+\theta x_{e2}-y_{e1}-y_{e2})}
\times
\]
\beq
\langle s_1, \nu_1, s_2, \nu_2 \vert s, \nu \rangle d^4 p_e
{\cal S}_2^{s_1 s_2:s} ({1 \over 2} (y_{e1}-y_{e2} ), p_e)
f( y_{e1}, s_1, \nu_1 , y_{e2},s_2 ,\nu_2 ).
\label{sc25}
\eeq
Using the fact that $\theta (x_{e1}-x_{e2}) \cdot  P_e = 
\left ((x_{e1}-x_{e2}) \cdot  P_e \right )^*$ gives
\[
= \sum_{\mu_1, \nu_1=-s_1}^{s_1} 
\sum_{\mu_2, \nu_2=-s_2}^{s_2}
\sum_{\mu, \nu=-s}^{s}
\int f^* ( x_{e1}, s_1, \mu_1 , x_{e2},s_2, \mu_2 ) 
({\cal S}_2^{s_1 s_2:s} ({1 \over 2} ( (x_{e1}-x_{e2}),p_e)
\langle s_1, \mu_1, s_2, \mu_2 \vert s, \mu \rangle)^*
\times
\]
\[
{D^s_{\mu\nu} ( p_e \cdot \sigma_e) \over p_e^2 + \lambda^2}
e^{i p_e \cdot  (\theta x_{e1}+\theta x_{e2}-y_{e1}-y_{e2})}
\times
\]
\beq
\langle s_1, \nu_1, s_2, \nu_2 \vert s ,\nu \rangle d^4 p_e
{\cal S}_2^{s_1 s_2:s} ({1 \over 2} (y_1-y_2), p_e)
f(x_1, s_1, \nu_1 , x_2,s_2 ,\nu_2 ).
\label{sc26}
\eeq
Because the ${\cal S}_2^{s_1 s_2:s}$ are analytic in $p_e$ and the Wigner functions are polynomials in the components of $p_e$,
as in the single-particle case, after integrating
over test functions in $X_e$ and $Y_e$ satisfying the Euclidean time
support condition, the integral over $p_e^0$ can be computed using the
residue theorem.
The only contributing pole is at $p^0= -i \omega_\lambda (\mathbf{p})$:
\[
=\int  \sum_{\mu_1, \nu_1=-s_1}^{s_1} 
\sum_{\mu_2, \nu_2=-s_2}^{s_2}
\sum_{\mu, \nu=-s}^{s}
\left ( \int dx_{e1} dx_{e2} f( (x_{e1}, s_1, \mu_1 , x_{e2},s_2 ,\mu_2 )\right .
\times
\]
\[
\left .
{\cal S}_2^{s_1 s_2:s} ({1 \over 2} (x_{e1}-x_{e2}), p_e)
\langle s_1, \mu_1, s_2, \mu_2 \vert s, \mu \rangle) 
e^{-\omega_\lambda (\mathbf{p})(x_{e1}^0+x_{e2}^0)}\right )^*
{\pi d\mathbf{p} \over \omega_\lambda (\mathbf{p}) } D^s_{\mu\nu} ( p_m \cdot \sigma_m)
\times
\]
\beq
\left ( \int dy_{e1} dy_{e2}
f( y_{e1}, s_1, \nu_1 , y_{e2} ,s_2 ,\nu_2 )
{\cal S}_2^{s_1 s_2:s} ({1 \over 2} (y_1-y_2), p_e)
\langle s_1, \nu_1, s_2, \nu_2 \vert s, \nu, \rangle
e^{-\omega_\lambda (\mathbf{p})(y_{e1}^0+y_{e2}^0)} \right ) 
\label{sc27}
\eeq
which is non-negative since it has the form $\sum_{mn} a_m^* P_{mn} a_n$
where $P_{mn}$ is positive definite, so it is positive.

Note that unlike the local case, the Euclidean time differences
$x^0_1-x^0_2$ do not have to be different from zero.  This is because
in the non-local case there are different $N$-point quasi Schwinger
functions with different numbers of initial and final coordinates.

This four-point example is easy to generalize;
\begin{itemize}
  
\item[1.] The connected four point function
  can be replaced by a square matrix of $m+n$-point connected
  distributions with the same ``intermediate states''.

\item[2.] The spinors of rank $(s,0)$ can be replaced by
  spinors of rank $(s_r,s_l)$ or direct sums of rank
  $(s,0)\oplus(0,s)$.

\item[3.] A single ``intermediate state'' of mass, $\lambda>0$, and spin $(s_r,s_l)$ can be
replaced by
linear superpositions of states with different $\lambda$'s and spins
with a positive weight.. 
  
\item[4.] The initial and final distributions
$
{\cal S}_2^{s_1 s_2:s} ({1 \over 2} (y_1-y_2), p_e)
\langle s_1, \nu_1, s_2, \nu_2 \vert s, \nu, \rangle$ 
are replaced sums of products of invariant distributions
and constant coupling coefficients of the form
\beq
\sum_a {\cal S}_{n,a}^{(s_{r1},s_{l1}) \cdots (s_{rn},s_{ln} ); (s_r, s_l) }
(X_e-x_{e1} \cdots X_e-x_{en-1} ; p^c_e)
C^{(s_{r1},s_{l1}), \cdots (s_{rn},s_{ln} ); (s_r s_l) }_{(\mu_{r1},\mu_{l1}) \cdots (\mu_{rn},\mu_{ln} ); (\mu_r, \mu_l) }(a) 
\label{sc28}
\eeq
where $X={1 \over n}(x_{e1} + \cdots + x_{en})$, each ${\cal S}_{n,a}$
is a connected is a Euclidean invaraint function of the $X_e-x_{ei}$ and $p_e$,
analytic in $p_e$, and the
coefficients $C^{(s_{r1},s_{l1}), \cdots (s_{rn},s_{ln} ); (s_r s_l) }_{(\mu_{r1},\mu_{l1} \cdots (\mu_{rn},\mu_{ln} ); (\mu_r, \mu_l) }(a) $ decompose the
tensor products of $D(A)$, $D(B)$ to direct sums with spin
$(s_r s_l)$, similar to (\ref{sc23}),(\ref{ir.13}).

It isstraightforward to construct ${S}_{n,a}$ with these properties.  In the
general case a single value of $m$ is replaced by a linear superpostion  
of states with different values of $m$ with a positive weight, $\rho (m)$:
\beq
m \to \int \rho(m) dm 
\eeq
  
\end{itemize}
The combinations (\ref{sc28}) should be symmetric or antisymmetric with respect to exchange of identical particles. This can be realized by projecting the
initial and final states on the symmetric or antisymmetric subspace of the Hilbert space.

With these generalizations the proof of Euclidean covariance and reflection
positivity follows the proof in the four-point case.

The discussion in this section shows that it is not difficult to
construct connected quasi-Schwinger functions that satisfy Euclidean
covariance and reflection positivity.  These can be used to
construct a Hilbert space inner product where the vectors are
function of Euclidean variables with support for positive Euclidean times.

\section{Cluster expansions}\label{sec8}

The motivation for exploring the Euclidean approach to relativistic
quantum mechanics is the difficulty in satisfying cluster properties
in relativistic direct interaction models.

In this section a generalization of the linked cluster theorem
is used to construct quasi Schwinger functions that satisfy
cluster properties using the connected quasi-Schwinger functions
introduced in the previous section.

In this section reflection positive quasi-Schwinger functions will be
expressed as linear combinations of products of connected
reflection positive quasi-Schwinger functions.  The connected
reflection positive quasi-Schwinger functions are the elementary
building blocks of reflection positive quasi-Schwinger functions that
satisfy cluster properties.  For systems of identical particles
the sums have to include all combinations of tensor products of
connected quasi-Schwinger functions that are generated by permutations.

Let  
\beq
\{ S_{mn}(x_{me}, \cdots, x_{1e}: y_{1e}, \cdots, y_{ne}) \}
\label{ip:1}
\eeq
be a collection reflection-positive quasi-Schwinger distribution $(1 \leq m,n \leq N \leq\infty$ for a system of identical particles.  In this and the following expressions the
spinor indices are suppressed.

Each $S_{mn}$ can be expanded as a sum of
tensor products of connected reflection positive kernels.
The tensor products contributing to this sum 
are products of a total of $l$ kernels;
$k_1$ connected kernels of type
$S^c_{m_1 n_1}$, $k_2$ connected kernels of the type
$S^c_{m_2 n_2}$, $\cdots$,  $k_l$ connected kernels of the type
$S^c_{m_l n_l}$ 
where 
\beq
n = \sum_{i=1}^l k_i n_i \qquad
m = \sum_{i=1}^l k_i m_i .
\label{ip:2}
\eeq
For systems of identical particles the sums include all distributions
generated by $m!$ permutations of the final
coordinates and $n!$ permutations of the initial coordinates.  

Assume each kernel, $S^c_{m_i n_i}$, is invariant up to sign under $m_i!n_i!$
permutations that separately interchange the initial and final arguments.
If a given kernel appears $k_i$ times in the product,
there are $k_i!$ additional permutations that exchange the $k_i$
identical terms in the product.  After accounting for these
invariances there remain
\beq
N= {m! n! \over k_1! \cdots k_l! (n_1!m_1!)^{k_1}
(n_2!m_2!)^{k_2} \cdots  (n_l!m_l!)^{k_l}}
\label{ip:3}
\eeq
kernels with this structure that differ by permutations, where the
integers in (\ref{ip:3}) are constrained by (\ref{ip:2}).
For identical particles for each product that contributes to 
$S_{mn}$, the sum must also include all $N$ distributions that
are generated by these additional permutations. 

It is possible to construct a generating function for the
$S_{mn}$ in terms of the individual connected $S^c_{mn}$. To do this consistently
for Bosons and Fermions define formal creation and annihilation operators.
The $a^{\dagger}(y_i)$ operators create initial states and
$\mathbf{b}^{\dagger}(x_i)$ create final states.  These operators satisfy 
\beq
[a (x_i), a^{\dagger}(y_i)]_{\pm}
= \delta^4(x_i-y_i)\delta_{\mu_{ir}\mu'_{ir}} \delta_{\mu_{il}\mu_{il}'}   
\label{ip:8}
\eeq
\beq
[b (x_i), b^{\dagger}(y_i)]_{\pm} = \delta^4(x_i-y_i)\delta_{\mu_{ir}\mu'_{ir} }\delta_{\mu_{il}\mu_{il}'}
\label{ip:9}
\eeq
\beq
[a (x_i), b (y_i)]_{\pm}=0
\label{ip:10}
\eeq
\beq
a (x_i) \vert 0 \rangle = b(x_i) \vert 0 \rangle =0 \qquad
\langle 0 \vert 0 \rangle =1. 
\label{ip:11}
\eeq
In these expressions, $[A,B]_-$ is a commutator and $[A,B]_+$ is an anticommutator.
The operators $a(x_i)$ and $b (y_i)$  are just formal operators that are useful for bookkeeping
purposes.  The same is true for the formal ``vacuum'', $\vert 0
\rangle$; it has nothing to do with the ground state of the theory.  They are
just for the purpose of constructing quasi-Schwinger functions from
the connected distributions.

With this notation define the generating functional, $S$, as the formal sum
\[
S = \sum_{mn} \frac{1}{m!n!} \int  d^{4m}xd^{4n}y  
\int b^{\dagger}(x_{me})  \cdots b^{\dagger}(x_{1e}) \times
\]
\beq
S_{mn}(x_{me},\cdots , x_{1e}:y_{1e}, \cdots y_{ne})    
a^{\dagger}(x_{1e})  \cdots a^{\dagger}(x_{ne}).
\label{ip:12}
\eeq
The individual symmetrized $S_{mn}(X:Y)$ can be extracted from $S$ using
products of annihilation operators and the formal vacuum vector
\beq
S_{mn}(X:Y)=
\langle 0 \vert 
a(y_{ne})  \cdots a(y_{1e})
b(x_{1e})  \cdots b(x_{me}) S \vert 0  \rangle .
\label{ip:13}
\eeq
Note that in (\ref{ip:13}) there are $n!m!$ pairings of
creation and annihilation
operators of the same type that are equivalent to
adding all possible exchanges.  If $S_{mn}$ is already symmetric or
antisymmetric each product of pairings gives the same result.  This
results in an  overcounting by $n!m!$ which is canceled
by the denominator
in (\ref{ip:12}).

Each kernel $S_{mn}$ is a sum of all allowed products of
connected kernels that
satisfy (\ref{ip:2}): 
\beq
S_{mn}= \sum  \prod_{i=1}^l (S^c_{m_i n_i})^{k_i}
\label{ip:14}
\eeq
where the sum is over all products of connected kernels satisfying
(\ref{ip:2}).  It is useful to express each one of these as
operators in terms of the creation and annihilation operators:
\[
S^c_i = \sum \frac{1}{m_i!n_i!} \int  d^{4m}x_{e}d^{4n}y_{e}  
\int b^{\dagger}(x_{me})  \cdots b^{\dagger}(x_{1e})
\]
\beq
S^c_{m_in_i}(x_{m_ie}\cdots x_{1_ie}:y_{1_ie} \cdots y_{n_ie})    
a^{\dagger}(y_{1_ie})  \cdots a^{\dagger}(y_{n_ie})
\label{ip:15}
\eeq
where the $S^c_{m_in_i}$ are connected factors that appear in the
product.

In this notation the operator $S$ is a sum of products of these
different types of connected operators.  While the factor
$\frac{1}{m_i!n_i!}$ divides by the number of permutations
give the same kernel when $S^c_{m_in_i}$ is symmetric, if a
given $S^c_i$ appears $k_i$ times in the product, there are
$k_i!$ permutations that exchange all of the coordinates
in each factor in the product.  In terms of the different
types of connected operators the generating function has
the form
\beq
S = \sum \frac{1}{k_1!} (S^c_1)^{k_1}  \cdots  \frac{l}{k_l!} (S^c_l)^{k_l}
=
e^{\sum_l S^c_l}
\label{ip:16}
\eeq
and the individual $S_{mn}$ can be extracted from this expression using
(\ref{ip:13}) which can be expressed as  
\beq
S_{mn}(X:Y)=
\langle 0 \vert 
a(y_{en})  \cdots a(y_{e1})
b(x_{e1})  \cdots b(x_{em}) e^{\sum_l S^c_l} \vert 0  \rangle .
\label{ip:17}
\eeq
which is the form of the linked cluster theorem for the
quasi-Schwinger distributions.  The convergence of the series is
irrelevant since only a finite number of terms contribute to a given
$S_{mn}(X:Y)$.

Note that while this generating function involves systems of arbitrary
numbers of degrees of freedom, it is possible the individual connected
components can involve a finite number of degrees of freedom. 

\section{Conclusion}

The purpose of this work is to show that by relaxing the requirement
of locality it is possible to construct a set of reflection positive
Euclidean covariant distributions satisfying cluster properties.  The
new feature is that a single N-point Schwinger function is replaced by
$N-1$ distributions with $m$ final and $k$ initial degrees of freedom
with $m+k=N$.  This simplifies the reflection positivity requirement.
These Euclidean distributions define the kernel of a Hilbert space
inner product.  There is a representation of the Poincar\'e Lie
algebra on this Hilbert space, \cite{gohin}.  These generators are self-adjoint
operators satisfying cluster properties.  The spectrum of the
Hamiltonian is bounded from below.  It follows
that there is a unitary representation of the Poincar\'e group that
satisfies space-like cluster properties and a spectral condition.
Given these distributions it is possible to perform any kind of
quantum calculation without the need for analytic continuation.

While establishing the existence of a large class of reflection
positive quasi Schwinger functions is an important first step for
constructing dynamical models, the problem is that this construction
assumed an acceptable spectral density, which is dynamical information
that should be calculated rather than assumed.

The formulation of a dynamical principle that could generate these
distributions is beyond the scope of this paper, but it in important
question that needs to be addressed in the future.

Some aspects of this program are discussed elsewhere.  They all assume
the existence of a set of distributions with the properties shown in
this work.  Reference \cite{Kopp:2011vv} provides a computational
justification that this formalism may be applied to compute scattering
cross sections.  The scattering computations used a variation of the
time-dependent formulation of scattering that utilized the invariance
principle \cite{simon} with narrow wave packets to approximate
sharp-momentum transition matrix elements.  Reference
\cite{Aiello:2015jgc} provided a formulation of scattering with
composite particles, using a generalization of Haag-Ruelle scattering.
This required a Euclidean construction to isolate composite one-body
states using functions of the mass operator, which is represented by
the Euclidean Laplacian, with compact support.  This was done by
establishing the completeness of polynomials in the Euclidean
Laplacian using the Carleman condition \cite{Carleman:1926}.
Reference \cite{gohin} provides explicit expressions for the
Poincar\'e generators with any spin and proved their self-adjointness .

One of the interesting observations about this approach is that
given reflection positive Euclidean distributions,  it is possible
to perform quantum mechanical calculations directly in a Euclidean
representation, without analytic continuation.

\bibliography{master_bibfile.bib}{}
\end{document}